# Insulator-to-superconductor transition in quasi-one-dimensional $HfS_3$ under pressure

Binbin Yue[1]*, Wei Zhong[1], Wen Deng[1], Ting Wen[1], Yonggang Wang[1], Yunyu Yin[2], Pengfei Shan[2,3], Xiaohui Yu[2,3,4]*, Fang Hong[2,3,4]*

[1] *Center for High Pressure Science & Technology Advanced Research, 10 East Xibeiwang Road, Haidian, Beijing 100094, China*

[2] *Beijing National Laboratory for Condensed Matter Physics, Institute of Physics, Chinese Academy of Sciences, Beijing 100190, China*

[3] *School of Physical Sciences, University of Chinese Academy of Sciences, Beijing 100190, China*

[4] *Songshan Lake Materials Laboratory, Dongguan, Guangdong 523808, China*

*Email: yuebb@hpstar.ac.cn; yuxh@iphy.ac.cn; hongfang@iphy.ac.cn

**Abstract**

Various transition metal trichalcogenides (TMTC) show the charge-density-wave and superconductivity, which provide an ideal platform to study the correlation between these two orderings and the mechanism of superconductivity. Currently, almost all metallic TMTC compounds can show superconductivity either at ambient pressure or at high pressure. However, most TMTC compounds are semiconductors and even insulators. Does the superconductivity exist in any non-metal TMTC compound? In this work, we managed to manipulate the electronic behavior of highly insulating $HfS_3$ in term of pressure. $HfS_3$ underwent an insulator-semiconductor transition near 17 GPa with a band gap reduce of ~1 eV. The optical absorption and Raman measurement provide the consistent results, suggesting the structural origin of the electronic transition. Upon further compression, $HfS_3$ becomes a superconductor. The superconducting transition was initialized as early as 50.6 GPa and the zero-resistance is reached above 91.2 GPa. The superconducting behavior is further confirmed by both the magnetic field effect and current effect. This work sheds the light that all TMTC may be superconductors, and opens a new avenue to explore the abundant emergence phenomena in TMTC material family.

## Introduction

Transition metal chalcogenides (TMCs) are a sort of layer-structured semiconductors or metals with great potential as next generation electronics and optoelectronics (1-3). There are two important branches in TMC material family, transition metal dichalcogenides (TMDCs) and transition metal trichalcogenides (TMTCs). Different from the TMDC in form of typical two-dimension structure, TMTCs show the characters of quasi-one-dimension properties. The chemical formula of TMTCs is $MX_3$, where M is transition metal atom belong to either group IVB (Ti, Zr, Hf) or VB (Nb, Ta) and X is chalcogen atom from group VIA (S, Se, Te). As shown in Table I, most of them crystalize in monoclinic structure ($P2_1/m$) while $NbSe_3$ and $TaSe_3$ can also stabilize in triclinic and orthorhombic structure, respectively (3, 4). Physical properties of $MX_3$ vary from superconducting (SC) to semiconducting (even highly insulating) according to the distinctive stacking sequences of the one-dimensional (1D) chain variants, which allows their application in various areas. Charge density waves (CDW) transitions have shown in $NbS_3$ (5, 6), $TaS_3$ (7-9), $NbSe_3$ (10-12) while coexistence with SC has also been observed in both $ZrTe_3$ (13, 14) and $HfTe_3$ (15). $TaSe_3$ is a semimetal which was proved to be a topological insulator recently (16, 17) and has a superconducting transition at ~2.2 K (18) as well. Trisulfides with M from group IVB are all semiconductors with band gap from ~1 to ~ 3 eV (3, 19-27). With the introduction of pressure, which can manipulate the layer distance and interaction in a large range, some TMC materials show various emergence phenomena. In $ZrTe_3$, $T_{CDW}$ increases to ~105 K at 1.1 GPa, while the filamentary SC disappears. $T_{CDW}$ shows a dome-like behavior and disappear at ~ 5 GPa when a bulk SC emerges (28-30). Suppression of the CDW and emerging of SC has also observed in metallic $NbSe_3$ (10-12) *o*-$TaS_3$ (31) under pressure, which is also observed in some TMDC materials, such as $TaS_2$ (32). It is noted that all these TMTC materials with SC transition are metallic at ambient pressure. Theoretical calculation shows semiconducting $TiS_3$ could be a superconductor at 80 GPa (33). However, it is still an open question that whether any semiconducting or insulating $MX_3$ could become a superconductor upon compression by experiment.

Among all $MX_3$ compounds, the band gap of $HfS_3$ is almost the largest (1.7~3.1 eV) which makes it highly insulating (23-27). Here we choose this material as a prototype to study the pressure effects on its electronic properties, with the aim to find out whether SC state in such an insulating $MX_3$ can be induced by pressure or not. In this work, we found that the band gap of $HfS_3$ is sensitive to external pressure. $HfS_3$ underwent an insulator-semiconductor transition near 17 GPa with sudden reduce of band gap. Upon further compression, signature of SC transition starts to show above 50.6 GPa and zero resistance is observed above 91.2 GPa. $T_c$ increases with further compression and reaches 8.1 K at 121 GPa. This work has demonstrated that the electronic properties of layered and insulating TMTC $HfS_3$ can be well manipulated by pressure, while the pressure plays a role similar with electron doping, and the SC transition is firstly achieved in such an insulating TMTC compound, which suggests other semiconducting TMTC are also promising superconductors under high pressure.

TABLE I Electronic properties of the electronic properties of $MX_3$ compounds

| $MX_3$ | | Chalcogen | | | | | | | | | | | |
|---|---|---|---|---|---|---|---|---|---|---|---|---|---|
| Group | M | S | | | | Se | | | | Te | | | |
| | | structure | Band gap (eV) | CDW | SC | structure | Band gap (eV) | CDW | SC | structure | Band gap (eV) | CDW | SC |
| IV | Ti | **Monoclinic $P2_1/m$**<br>$P2_1/m$-to-$P2_1/m$ at 22 GPa<br>(3, 19, 20, 35-37) | **0.8-1.1** | **29 K, 53 K, 103 K** | - | No reported synthesis | | | | **$P2_1/m$ *** (34) | **0.69** | - | - |
| IV | Zr | **$P2_1/m$** (20-23, 38) | **1.9-2.8** | - | - | **$P2_1/m$** (20-22) | **1.1-1.3** | - | - | **$P2_1/m$** (13, 28-30, 39) | **metal** | **63 K,** to 105 K at 1.1 GPa, disappears above 5 GPa. | **2 K** suppressed at 1.1 GPa; reappear at 2.5 K and 5GPa |
| IV | Hf | **$P2_1/m$** (23-27) | **1.7-3.1** | - | - | **$P2_1/m$** (22) | **1.0** | - | - | **$P2_1/m$** (15) | **metal** | **82** to 99 K at 1 GPa | **1.4** |
| V | Nb | **Triclinic $P$-1** (40) | **0.83** | - | - | **$P2_1/m$** (10-12) | **metal** | **$T_1$: 145 K, $T_2$: 59 K** both suppressed by pressure | 5 K at 4 GPa | **$P2_1/m$ *** (34) | **0.486** | - | - |
| V | Nb | **$P2_1/m$** (5, 6) | **metal** | **450-475 K, 360 K, 150 K** | - | | | | | | | | |
| V | Ta | **Orthorhombic** (7-9, 31, 41) | **metal** | **210-220 K** disappear at 11.5 GPa | 3.1 K at 11.5 GPa | **$P2_1/m$** (16-18) | **metal & TI** | - | **~2.2** | No reported synthesis | | | |
| V | Ta | **$P2_1/m$** (7) | **metal** | **240 K, 160 K** | - | | | | | | | | |

* Newly synthesized in single-chain form inside carbon nanotube, not stable in bulk (34).
CDW: charge density wave; SC: superconducting; TI: topological insulator
Note: Metallic $NbS_3$ has been obtained by heating $NbS_3$ semiconductor and is reported to be superconducting at 2.15 K (42). Other polymorphous have also been reported for $NbS_3$ (4, 43).

## Experiments:

The standard four-probe electrical resistance measurement under high pressure was measured in a commercial cryostat from 1.7 K to 300 K by a Keithley 6221 current source and a 2182A nanovoltmeter. A BeCu alloy diamond anvil cell (DAC) with two opposing anvils was used to generate high pressure. For Run #1, 300 μm culets were used and 100 μm culets were used for Run #2. In these experiments, a thin single crystal sample was loaded into the sample chamber in a rhenium gasket with c-BN insulating layer. A ruby ball is loaded to serve as in-ternal pressure standard for Run #1, while the Raman signal of diamond culet was used to calibrate the pressure for Run #2. The high-pressure Raman spectra were collected using a Renishaw Micro-Raman spectroscopy system equipped with a second-harmonic Nd:YAG laser (operating at 532 nm). A 532 nm laser was used, with a spot size of 1-2 μm. A pair of low fluorescent diamonds were used for Raman measurement. The laser power was maintained at relatively low power level to avoid overheating during measurements. KBr was used as pressure medium. In-situ high-pressure UV-VIS absorption spectroscopy was performed on a home-designed spectroscopy system (Ideaoptics, Shanghai, China) with a 405 nm laser. The high-pressure infrared experiments were performed at room temperature on a Bruker VERTEX 70v infrared spectroscopy system with HYPERION 2000 microscope. A thin $HfS_3$ single crystal was loaded and KBr was used as pressure medium. The spectra were collected in transmission mode in the range of 600-8000 $cm^{-1}$ with a resolution of 4 $cm^{-1}$ through a ~50×50 $\mu m^2$ aperture.

**Results and discussions:**

The electrical transport measurements on single crystal $HfS_3$ were conducted as a function of temperature up to 121 GPa. Two runs of experiments have been conducted and the resistance-temperature (*R*-*T*) curves are shown in Fig. 1(a) and 1(b), respectively. At ambient condition, the resistance of the $HfS_3$ sample is too large to be measured, the multimeter shows readable resistance value ~31 kΩ from 17.7 GPa at 300 K and ~8.8 MΩ at 2 K. As pressure further increases, the temperature dependent resistance always shows semiconductor feature while the resistance decreases monotonously with pressure, as shown in Fig. 1(a). However, staring from 50.6 GPa, the resistance below 5 K shows some anomaly as seen in inset of Fig. 1(a), and the phenomenon is much clearer at 65.1 GPa with a sudden drop of the resistance below ~5 K, as shown in Fig. 1(b). Such a sharp transition suggests a possible superconducting transition. This is further confirmed by the resistance measurement at much higher pressure. The sample reaches zero resistance at ~1.8 K at 91.2 GPa and further compression can move both the $T_c^{onset}$ and $T_c^{zero}$ to higher temperature (the $T_c^{\mathrm{onset}}$ is determined from the cross point of two straight lines above and below the transition, while $T_c^{\mathrm{zero}}$ is defined as zero-resistance temperature), as seen in the inset of Fig. 1(b). Another fact that can be noted is that above the superconducting temperature, the *R*-*T* curves still exhibit a semiconducting feature. From 91.2 GPa, the high temperature resistance becomes almost stable but still no typical metallic feature, different from the pressure induced SC transition behavior in other TMCs materials, like $MoS_2$ (44), $ReS_2$ (45), *o*-$TaS_3$ (31), *et al*. This phenomenon and the *R-T* trend looks quite similar with the SC transition observed in some semimetals (46-48). It is possible that $HfS_3$ enters a semimetal state instead of a good metallic state under pressure, since the resistance change with temperature is relatively small, compared with a typical semiconductor. The semiconducting behavior could origin from the non-hydrostatic condition under such a high pressure, which could induce both defect and grain boundary, and such effects usually cause a semimetal to show a semiconducting behavior. Meanwhile, we also note that the electronic transport behavior in $HfS_3$ under pressure is very similar with the electron doped $SrFBiS_2$ (49), in which La doping can induce the insulator-to-superconductor transition. In this case, the pressure effect in $HfS_3$ behaviors to some extent similar with the electron doping, and it is also a common sense that pressure (if it is high enough) generally increases the electron density of state near Fermi surface in most materials. It can be clearly seen that the $T_c^{\mathrm{onset}}$ increases with pressure with a small drop at 91.2 GPa, at which the superconducting transition is completed and reaches zero resistance. At the highest pressure of 121 GPa, $T_c^{\mathrm{onset}}$ reaches to 8.1 K with a $T_c^{\mathrm{zero}}$ of 2.8 K.

To further verify the pressure-induced superconductivity in $HfS_3$, electric transport measurements under external magnetic field and various currents have both been conducted. Fig. 2 (a) displays the R-T curve at 121 GPa under various magnetic fields up to 2 T. It can be seen clearly that $T_c$ is suppressed to lower temperature with increasing magnetic field, which is typical for a bulk superconducting transition. The upper critical field $\mu_0H_{c2}$ as a function of critical temperature $T_c^{\mathrm{onset}}$ is plotted in Fig. 2 (b) and the zero-temperature $\mu_0H_{c2}(0)$ is estimated to be 8.7 T by fitting the $\mu_0H_{c2}(T)$ with the Ginzburg-Landau (GL) equation (50). This value is lower than the Bardeen-Copper-Schrieffer (BCS) weak-coupling Pauli paramagnetic limit of $\mu_0H_p$ $=1.84T_c$ $=14.9$ T for $T_c = 8.1$ K, which implies a phonon-mediated superconductivity in $HfS_3$. In addition, the upward curvature of $\mu_0H_p$-$T$, similar with the phenomena observed in $ZrTe_3$ (39), $MgB_2$ (51) and several TMDCs (44, 52, 53), could be a hint of possible multiband superconducting pairing state. By increasing the applied current, the normal state resistance remains unchanged. However, a broadening of the superconducting transition is observed together with the downward shift of both $T_c^{onset}$ and $T_c^{zero}$.

The resistance of $HfS_3$ crystal becomes measurable only from above 17.7 GPa, which suggests that the electronic transition might take place near this pressure point. To better understanding the electronic structure evolution, optical absorbance of $HfS_3$ as a function of pressure via both UV-VIS and IR spectroscopy has been carried out (Fig. 3 (a-c)). At ambient pressure, $HfS_3$ shows a band gap of 2.45 eV, consistence with previous results (24-26). As shown in Fig. 3 (d), the band gap decreases gradually with pressure and then has a sudden drop above 15.3 GPa, indicating a phase transition which is consistence with the electric transport results. Above 17.9 GPa, the band gap is out of the detecting range of UV-VIS spectroscopy. The color of the sample also changes from transparence to black (Fig. 3e), which makes the resistance of $HfS_3$ to be small enough to be detected by a multimeter. To get information under higher pressure, IR absorption measurements has been conducted, as displayed in Fig. 3(b-c) and results show that the absorbance is almost 0 up to 12.9 GPa, following with an increasing trend with pressure. Above 15.6 GPa, the absorbance drops a little bit due to the phase transition and then increases again. A band gap of 0.82 eV is obtained at 19 GPa, which is much smaller as compared with that of the ambient phase. Under further compression, it decreases slowly and reaches 0.36 eV at 53.7 GPa. The transmission spectra also agree with this. The transmittance decreases with pressure and was strongly suppressed at the highest pressure of 53.7 GPa. Meanwhile, the optical interference oscillation amplitude of the transmittance signal is also much smaller at 53.7 GPa. This suggested that the band gap $HfS_3$ is gradually closing and may started to transform to a semimetal state. The optical analysis confirms that there is an electronic transition near 17 GPa, the band gap is reduced by ~1 eV near this transition, which is consistent with the electric transport behavior in $HfS_3$. When pressure was released, the band gap was recovered with a small hysteresis. A band gap of 2.34 eV is obtained when the pressure was released to 0.4 GPa, close to that at ambient condition.

To understand the mechanism of the electronic transition observed by transport measurement and optical absorption experiment, we collected the Raman spectra under high pressure, as presented in Fig. 4(a). At ambient temperature and pressure, the crystal structure of $HfS_3$ belongs to the $C_{2h}$ point group and shows typical quasi-one-dimension feature, as seen in Fig. 4 (b-c). The Raman-active phonon modes can be represented by the irreducible representations as $\Gamma = 8A_g + 4B_g$ (54). $A_g$ phonon modes vibrate perpendicular to the quasi-one-dimension chain and are prominent while $B_g$ modes involving atomic displacement along the chain are weak. The Raman modes observed in this work can be divided into four groups. Group I have three $A_g$ lines at 74, 113 and 132 $cm^{-1}$ and one $B_g$ mode at 141 $cm^{-1}$ as shown in Fig. 4 (a) and 4 (d). It stems from rigid-chain modes and is associated with the out-of-plane vibration of 1D-like chains extending along the b-axis. The intensity of 74 $cm^{-1}$ peak is quite weak compared with others and cannot be detected above 7 GPa. All these three $A_g$ modes share the similar pressure coefficients within the range of 1.5 ~ 2 $cm^{-1}$/GPa. The $B_g$ mode at 141 $cm^{-1}$ shows a relatively higher $d\omega/dp$ value and starts to decay above 7 GPa. At around 16~18 GPa, all peaks in this group disappear while a new mode at 159 $cm^{-1}$ appears, indicating that a phase transition should occur. Group II corresponds to internal deformation of the chains and includes 3 $B_g$ modes (219, 246 and 260 $cm^{-1}$) and 2 $A_g$ modes (267 and 274 $cm^{-1}$). The first two $B_g$ modes show different pressure coefficients and merge together above ~7 GPa. It finally disappears above 13.6 GPa while another mode at 238 $cm^{-1}$ starts to show up. The $A_g$ mode at 274 $cm^{-1}$ shows a quite high value of pressure coefficient and becomes undetectable above 7 GPa. For 260 and 267 $cm^{-1}$ modes, a reversal of relative intensity can be observed at 3.1 GPa. Both of them last up to 19.7 GPa and then are replaced by a new mode at 321.4 $cm^{-1}$. Similar with group II, group III also corresponds to internal deformation of the chains. In this group, the $A_g$ line at 322 $cm^{-1}$ shows a moderate $d\omega/dp$ coefficient as compared with that of other modes. The other mode in this group, the $A_g$ line at 353 $cm^{-1}$, decays a lot from 3.1 GPa and then raise up from 13.6 GPa. The strengthening of this mode should be corresponding to the intermediate phase since it disappears above 19.7 GPa together with the 322 $cm^{-1}$, and then the 385 $cm^{-1}$ mode comes out, as shown in Fig. 4 (a) and 4 (e). Group IV has one strong $A_g$ line at 524 $cm^{-1}$, which is predominantly made of in-plane out of phase motion of $S_{II}$-$S_{III}$ pair (Fig. 4c). It shows a negative pressure coefficient for the original low-pressure phase and then a positive trend with the highest $d\omega/dp$ value than the other new modes (Table II) after the phase transition above 19.7 GPa. Based on the Raman results, we found that the electronic transition near 17 GPa is due to a structural phase transition, which started from ~13.6 GPa and completed at ~19.7 GPa. The strongly suppression of Raman signal above 40 GPa also suggests that $HfS_3$ could start to show some feature of a metal in which Raman signal is generally absent or very weak. According to previous study on $TiS_3$, there is an isostructural phase transition near 22 GPa (35), which may also apply to the phase transition near 17 GPa observed in $HfS_3$. Further study is still required to reveal the accurate structural information under high pressure.

TABLE II Pressure coefficients and Grüneisen parameters of Raman modes in the ambient phase and high-pressure phase.

| ambient phase | | | | HP phase | |
|---|---|---|---|---|---|
| Mode frequency ($cm^{-1}$) | symmetry | $d\omega/dp$ ($cm^{-1}$/GPa) | $\gamma$ | mode | $d\omega/dp$ ($cm^{-1}$/GPa) |
| 74 | Ag | 1.94 | 0.839 | | |
| 113 | Ag | 1.50 | 0.412 | | |
| 132 | Ag | 1.92 | 0.448 | $I_{hp}$ | 0.43 |
| 141 | Bg | 3.34 | 0.733 | | |
| 219 | Bg | 3.97 | 0.562 | $II_{hp}$ | 1.5 |
| 246 | Bg | 1.51 | 0.190 | | |
| 260 | Bg | 2.33 | 0.278 | | |
| 267 | Ag | 3.27 | 0.378 | $III_{hp}$ | 2.5 |
| 274 | Ag | 4.74 | 0.537 | | |
| 322 | Ag | 2.28 | 0.219 | $IV_{hp}$ | 2.7 |
| 353 | Ag | 3.35 | 0.294 | | |
| 524 | Ag | -0.21 | -0.012 | $V_{hp}$ | 2.9 |

Grüneisen parameters of all modes of the ambient phase are calculated by using the relationship $\gamma_i=(B_0/\omega_i)(d\omega_i/dp)$, where $B_0$ is the bulk modulus of $HfS_3$. All results are listed in Table II. The quite large value of $\gamma$ of the rigid $A_g$ mode 74 $cm^{-1}$ and $B_g$ mode 141 $cm^{-1}$ implies more contribution to the thermal expansion coefficient, similar as the situation in $TiS_3$ (37). Another phenomenon that needs to be noted is that the peak intensity of all high-pressure modes is much weaker than that of the ambient phase. It further decreases with pressure and mode $I_{hp}$ becomes undetectable at the highest pressure (40.5 GPa) reached in this experiment.

Based on spectroscopy and electric transport measurements, the pressure-temperature (*P-T*) phase diagram of $HfS_3$ has been plotted and is shown in Fig.5. At low pressure, $HfS_3$ is an insulator with an unmeasurable resistance and a band gap larger than 2 eV. From ~ 17.7 GPa, it enters a new semiconducting phase with a much lower resistance and smaller bandgap. The phase transition completes at ~ 20 GPa and the new phase lasts until 50.6 GPa. From this pressure point, superconductivity emerges from a semimetal state upon cooling. Within the highest-pressure limit (121 GPa) in this work, $T_c^{zero}$ increases with pressure, which differs from the decreasing trend of theoretically predicted $T_c$ in $TiS_3$ (33), but resembles that of $ZrTe_3$ (39).

**Conclusion**

In conclusion, $HfS_3$ undergoes an insulator-to-superconductor transition under high pressure up to 121 GPa. Near 17 GPa, a sudden drop of band gap from ~ 2 eV to ~ 0.6 eV is observed and then $HfS_3$ started to be conductive and measurable by multimeter, suggesting the insulator-semiconductor transition. The superconducting transition appears above 50.6 GPa while $HfS_3$ enters a semimetal state. Zero resistance is reached from 91.2 GPa and $T_c$ increase with pressure from 3.8 K at 65.1 GPa to 8.1 at 121 GPa. Our present finding represents the first experimental realization of pressure-induced superconductivity in a pristine insulating TMTC material with a record-high $T_c$ as well in the whole TMTC family.

**Author contribution**

B. B. Yue and F. Hong conceived the project. B. B. Yue, W. Deng, Y. Y. Yin, P. F. Shan, X. H. Yu and F. Hong did the transport measurement. B. B. Yue and W. Zhong loaded the sample for the optical measurement. B. B. Yue, W. Zhong, T. Wen, and Y. G. Wang did the UV-VIS absorption experiment. B. B. Yue and W. Zhong did the infrared absorption and Raman measurement. B. B. Yue and F. Hong wrote the manuscript. All authors made comments on the manuscript.

## Acknowledgment

This work was supported by the National Natural Science Foundation of China (Grant No. 12004014, U1930401) and Major Program of National Natural Science Foundation of China (22090041).

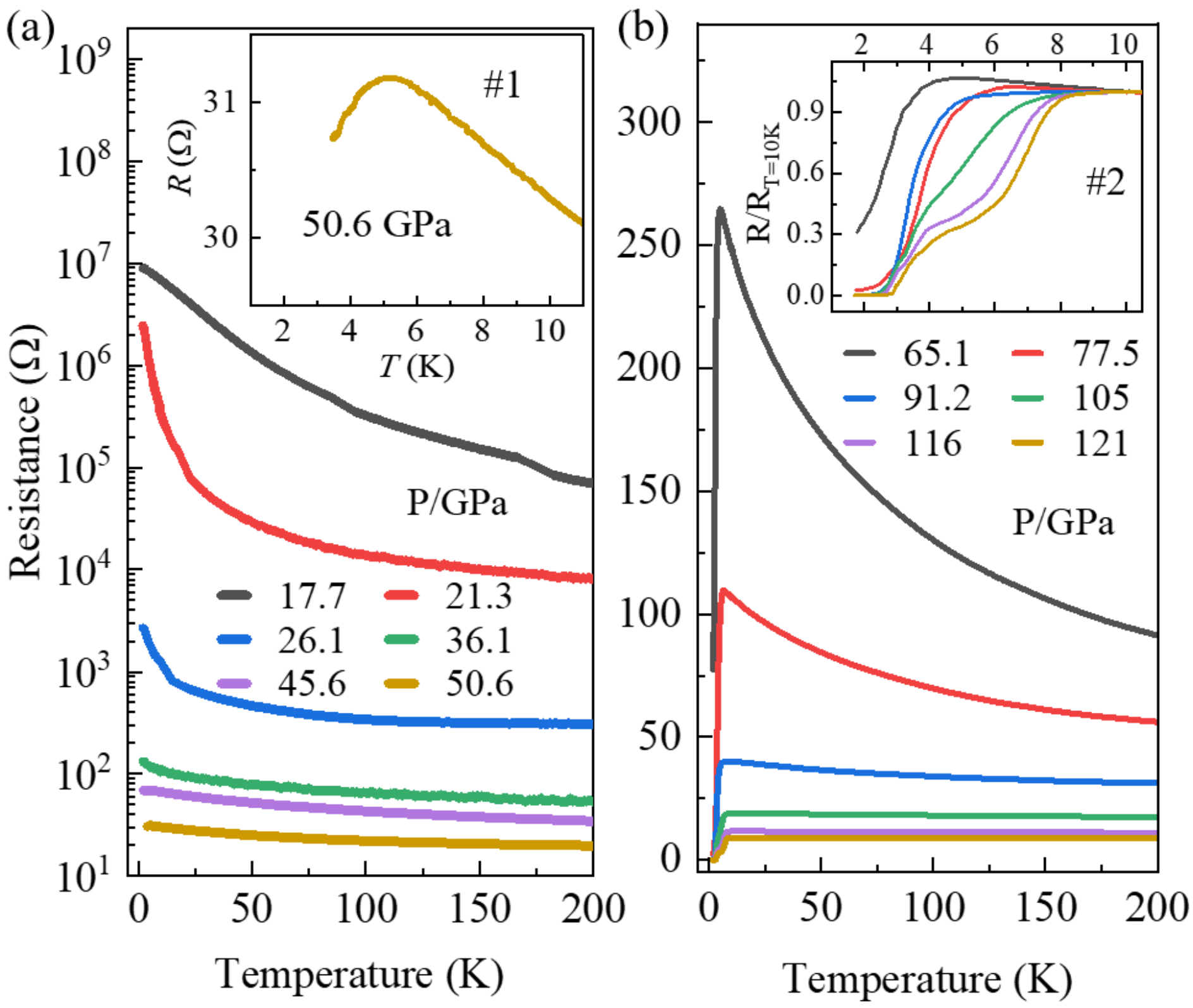

Figure 1 The electric transport properties at different pressures. (a) *R*-*T* curves for Run #1 up to 50.6 GPa, (b) *R*-*T* curves for Run #2 up to 121 GPa. Insets in (a) and (b) are enlarges of the low temperature range of *R*-*T* curves showing the superconducting transition.

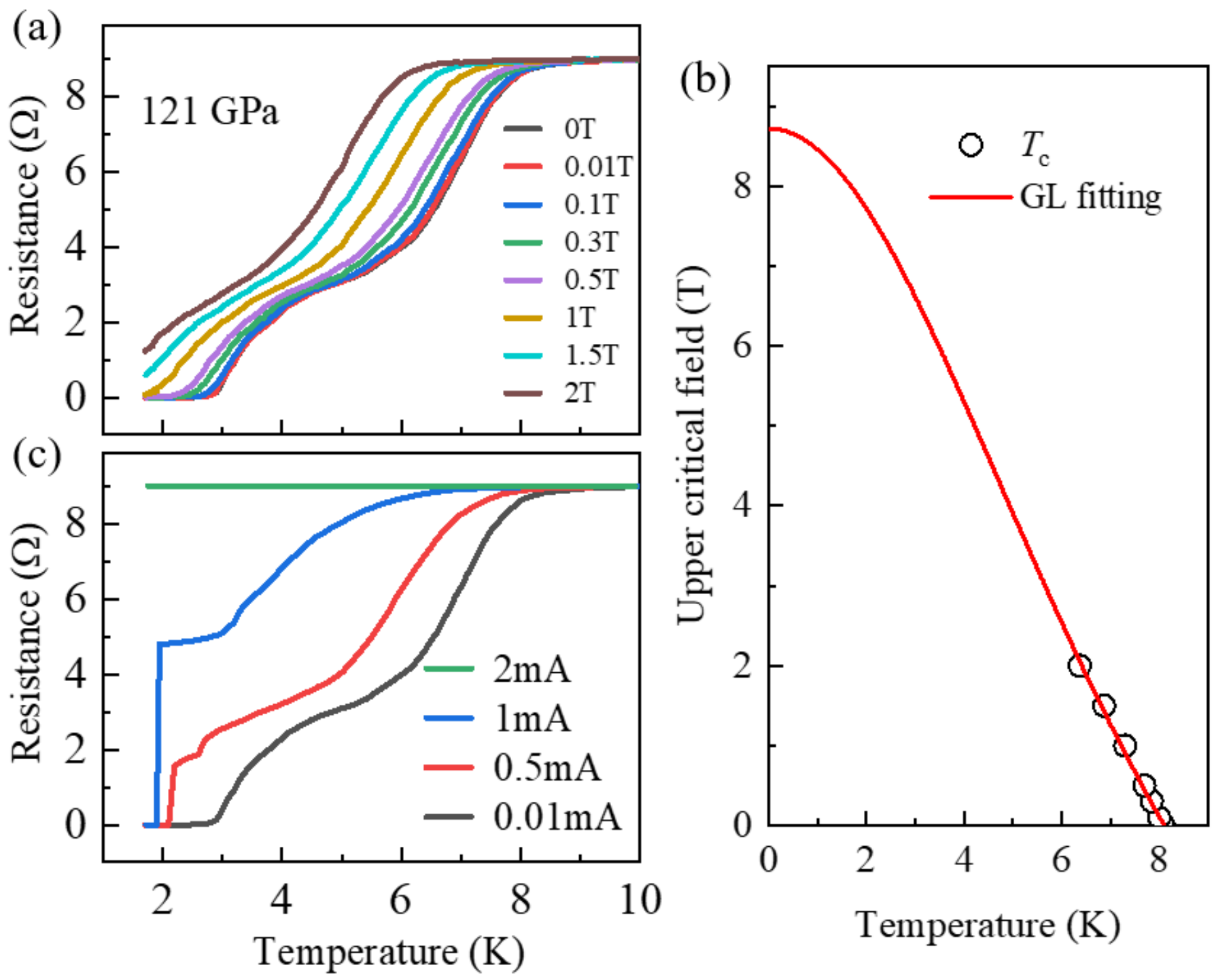

Figure 2 (a) Magnetic field effect on the superconducting transition of $HfS_3$ at 121 GPa, (b) upper critical magnetic field $\mu_0 H_{c2}$-$T$ phase diagram. The solid line in (b) represents G-L equation fitting. (c) Current effect on the superconducting transition of $HfS_3$ at 121 GPa.

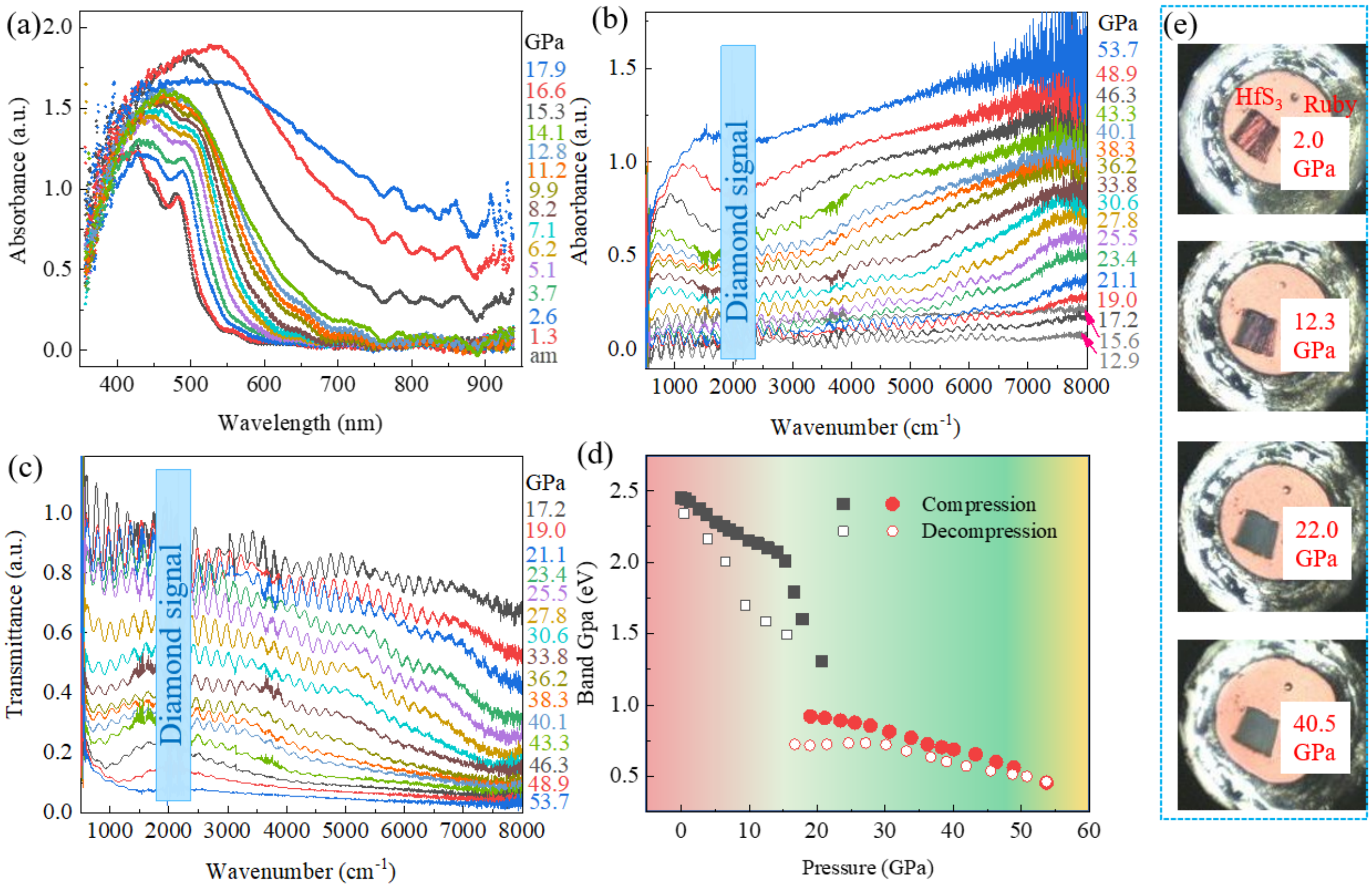

Figure 3 Optical and electronic properties of $HfS_3$ under pressure. (a-c) The UV-VIS absorption spectroscopy, infrared absorption spectroscopy and infrared transmission spectroscopy of $HfS_3$ under high pressure, respectively. (d) The pressure dependent optical band gap. (e) The optical image of samples at selected pressures inside diamond anvil cell.

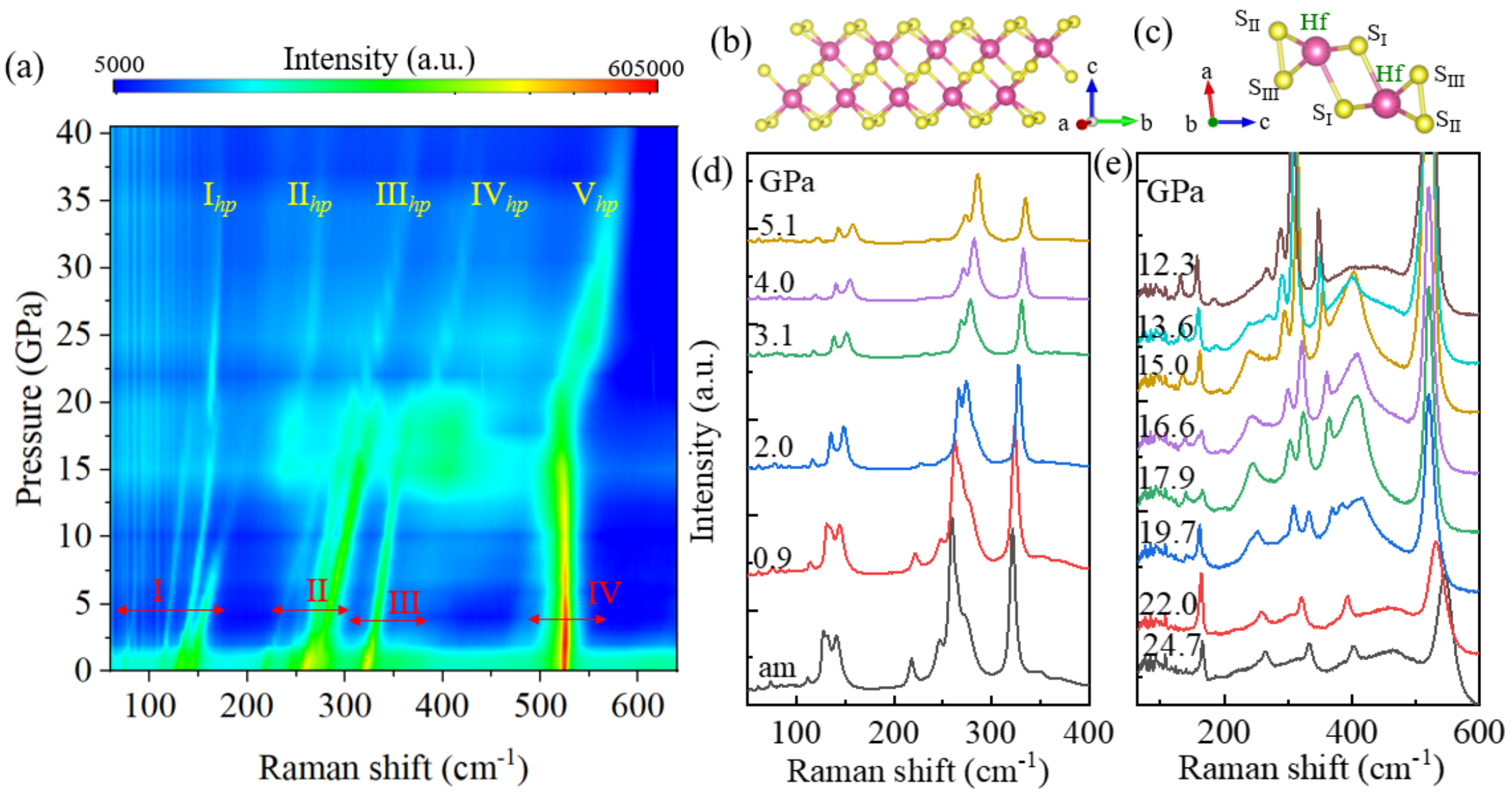

Figure 4 Pressure-dependent vibrational properties of $HfS_3$ investigated by *in situ* Raman spectroscopy. (a) 3D contour plot of Raman modes with pressure up to 40.5 GPa. (b) Schematic of the quisi-1D $HfS_3$ b-axis chains and (c) cross-section schematic view along the b axis with Hf atoms in pink and S atoms in yellow. (d & e) Selected Raman spectroscopy showing the critical change upon compression.

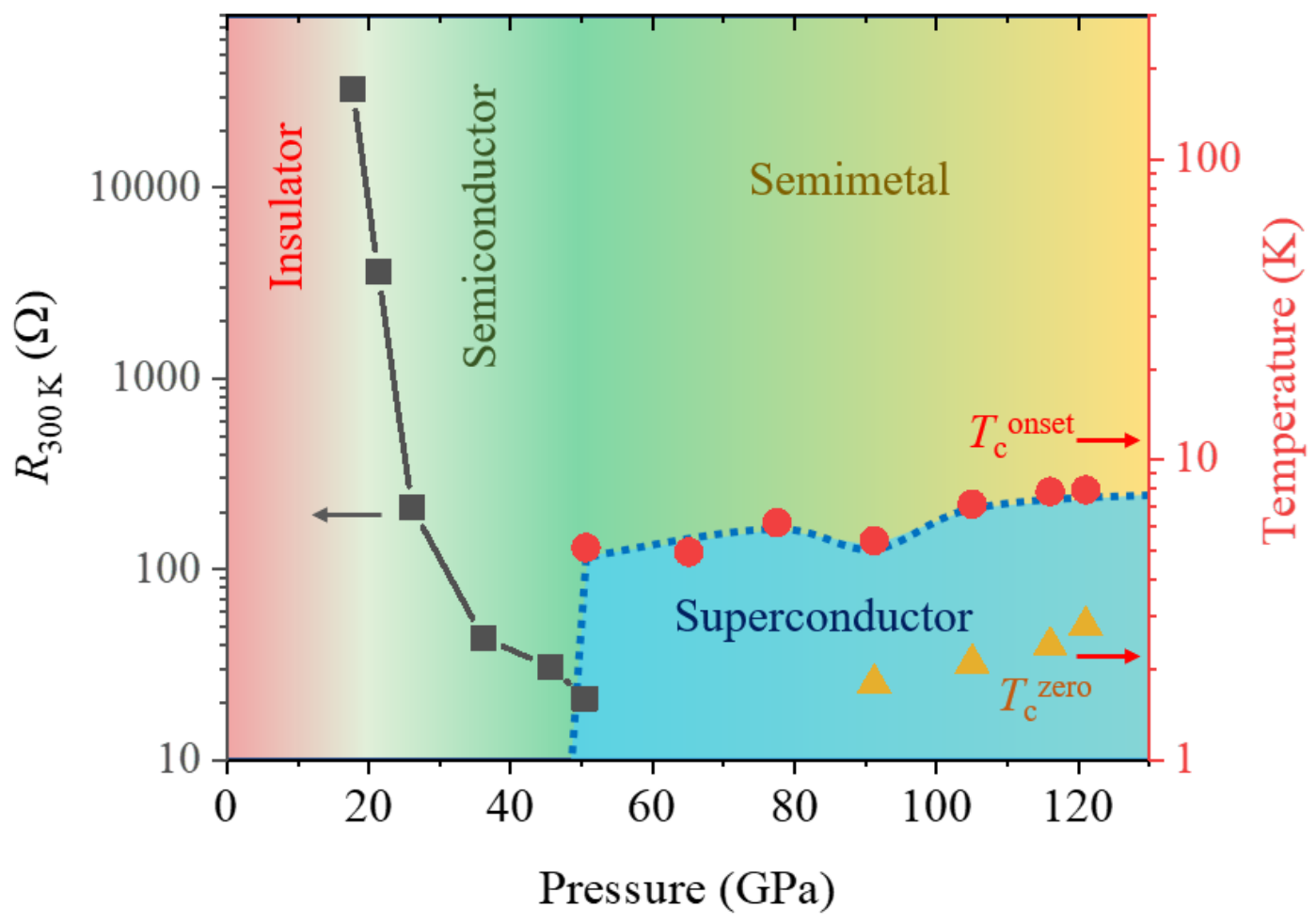

Figure 5 Pressure-temperature phase diagram of $HfS_3$. Black squares represent the resistance of $HfS_3$ at 300 K. Red circles and orange triangles show the $T_c^{onset}$ and $T_c^{zero}$, respectively. Dotted line is plotted to guide for eyes.

# Insulator-to-superconductor transition in quasi-one-dimensional $HfS_3$ under pressure

Binbin Yue[1]*, Wei Zhong[1], Wen Deng[1], Ting Wen[1], Yonggang Wang[1], Yunyu Yin[2], Pengfei Shan[2,3], Xiaohui Yu[2,3,4]*, Fang Hong[2,3,4]*

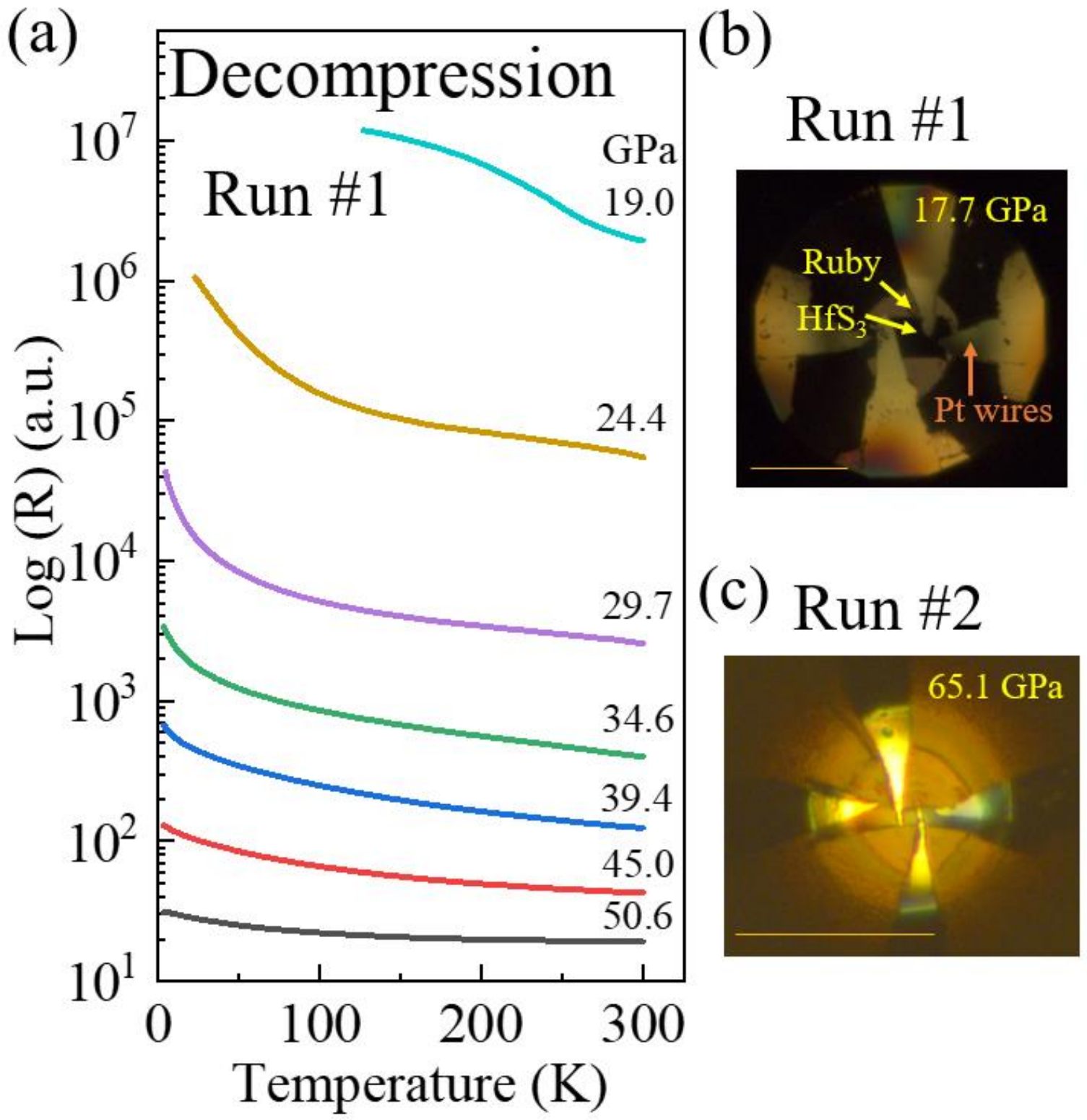

Figure S1 (a) R-T curves during decompression for Run #1. The resistance in low temperature range at 19.0 GPa is out of range, indicating that $HfS_3$ transfers back to insulating state during pressure release. Photos showing the standard four-probe electrical resistance measurement set up in the DAC chamber for Run #1(b) and #2(c). Scale bars in both (b) and (c) are 100 μm.

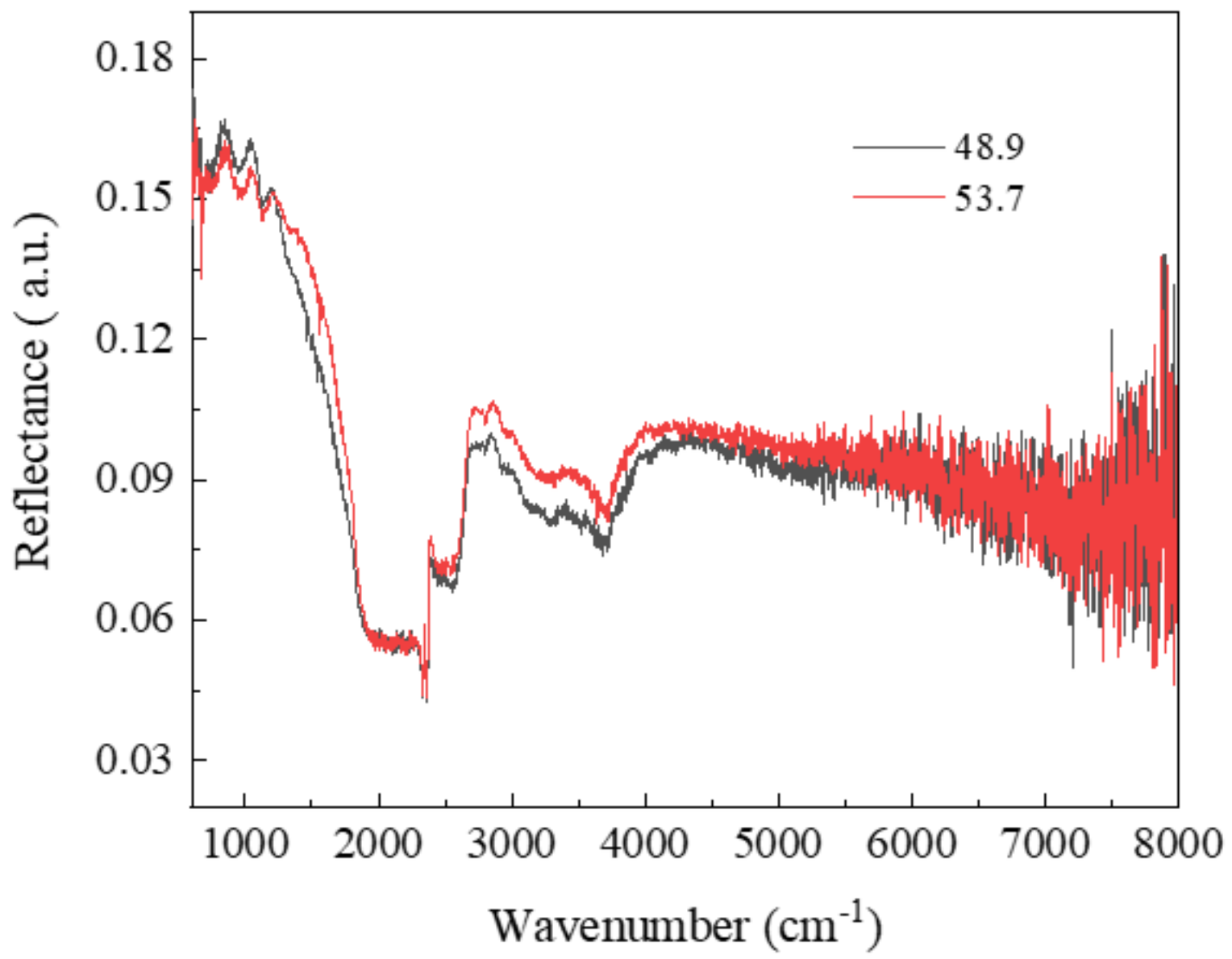

Figure S2 Reflectance spectra of $HfS_3$ at 48.9 and 53.7 GPa. The later shows slightly higher reflectance while none of them agree with the typical metallic behavior.

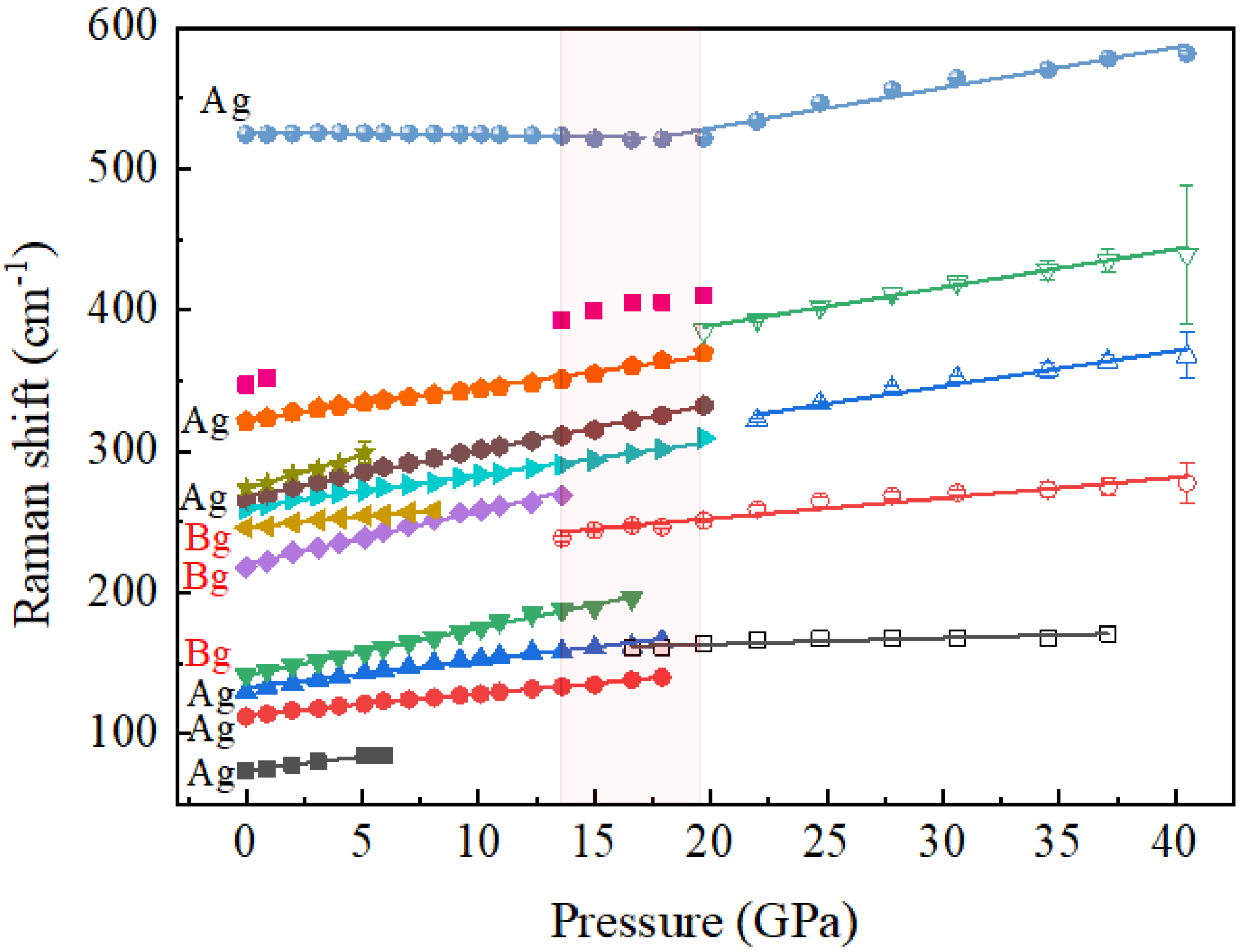

Figure S3 The pressure dependent Raman modes based on peak fitting results.

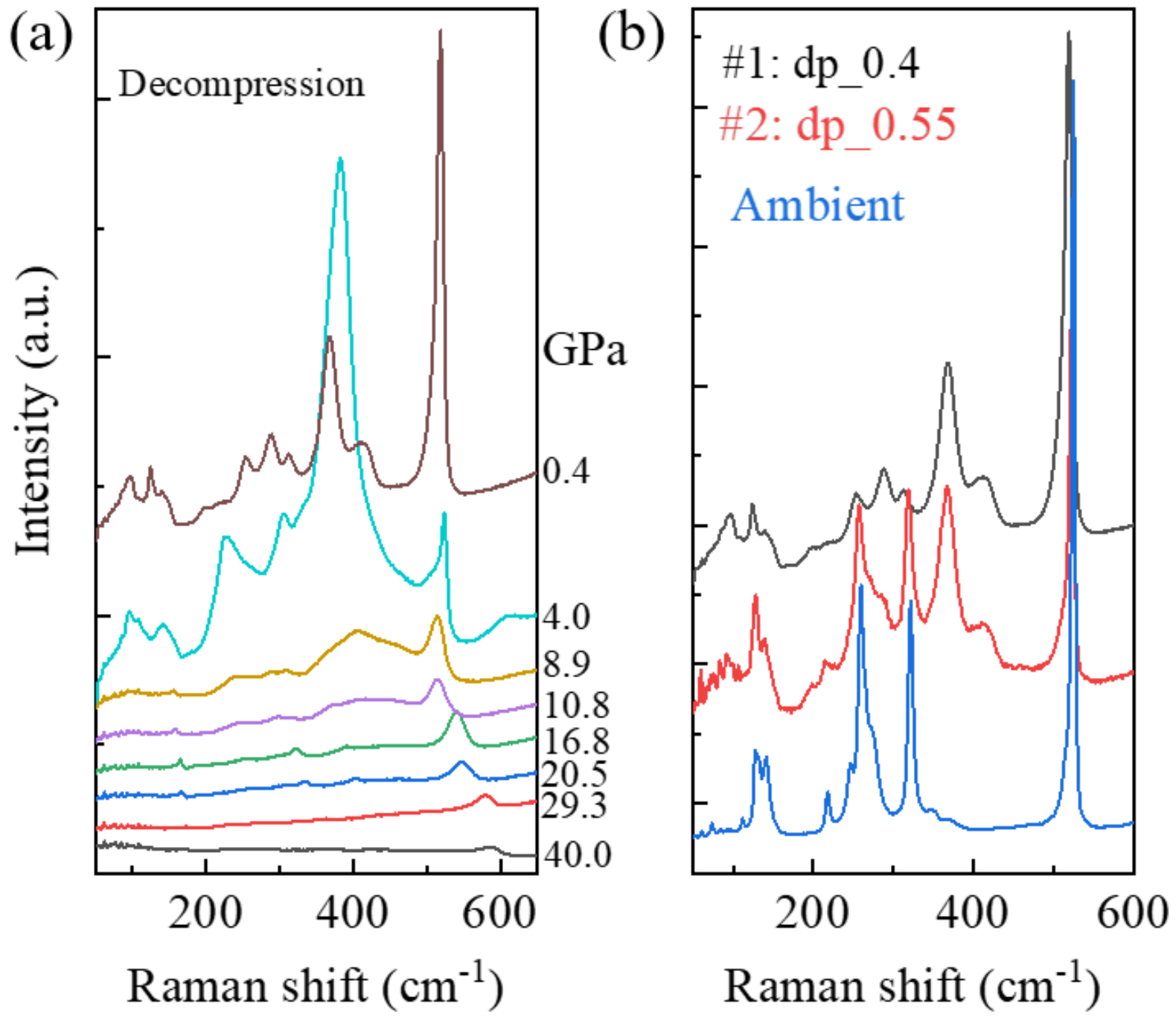

Figure S4 (a) Raman spectra during pressure release. (b) Raman spectra before and after high-pressure treatment.